\documentclass[11pt,dvips]{article}

\usepackage{rotating}
\usepackage{axodraw}
\usepackage{epsfig}
\usepackage{color}
\usepackage{hhline}
\usepackage{graphicx}
\usepackage{url}

\begin{document}

\begin{flushright}
  UCB-PTH-08/68\\
  IPMU 08-0064 \\
  KIAS-P08050 \\
  HD-THEP-08-23 \\[2mm]
  \today \\
\end{flushright}

\vspace{1cm}

\begin{center}
{\large \bf Determining Spin through Quantum Azimuthal-Angle
            Correlations}\\[1.5cm]
{ Matthew R. Buckley$^{1,2,3,4}$, Seong Youl Choi$^{5}$, Kentarou
  Mawatari$^{6,7}$ and Hitoshi Murayama$^{1,2,3}$}\\[1.cm] 
{\it $^1$ Department of Physics, University of California, Berkeley,
          CA 94720, USA\\[1mm]
     $^2$ Theoretical Physics Group, LBNL, Berkeley, CA 94720,
          USA\\[1mm] 
     $^3$ IPMU, University of Tokyo, 5-1-5 Kashiwa-no-ha, Kashiwa,
                Chiba 277-8568, Japan\\[1mm]
     $^4$ Address since 8/2008: California Institute of Technology,~Pasadena, CA 91125, USA \\[1mm]
     $^5$ Department of Physics and RIPC, Chonbuk National University,
          Jeonju 561-756, Korea\\[1mm]
     $^6$ School of Physics, Korea Institute for Advanced Study,
          Seoul 130-722, Korea\\[1mm]
     $^7$ Address since 10/2008: Institut f\"ur Theoretische Physik,
 Universit\"at Heidelberg, \\ Philosophenweg 16, D-69120 Heidelberg, Germany
}
\end{center}

\vspace{2cm}

\begin{abstract}
\noindent
Determining the spin of new particles is critical in identifying the
true theory among various extensions of the Standard Model at the next
generation of colliders.  Quantum interference between different
helicity amplitudes was shown to be effective when the final state is
fully reconstructible.  However, many interesting new physics
processes allow only for partial reconstruction.  In this paper, we
show how the interference effect can be unambiguously extracted even
in processes that have two-fold ambiguity, by considering the
correlation between two decay planes in $e^+ e^-$ collisions.\\[5mm] 
\noindent
PACS numbers: 12.60.-i, 12.60.Jv, 14.80.Ly
\end{abstract}

\newpage

The Large Hadron Collider (LHC) will soon usher us into the arena
of the electroweak
symmetry breaking scale and beyond.
Once the TeV regime is explored, it is highly anticipated that
the true theory for the origin and stability of the electroweak
scale \cite{Weinberg:1976} will be revealed.\\

One possible result from the LHC, generically predicted in new most models, is
the presence of new particles partnered with some or all of the SM particles.
For instance, every SM particle in the minimal supersymmetric
standard model (MSSM) \cite{Nilles, Drees} has a heavier partner whose spin
differs by 1/2. Alternatively, in the minimal universal extra
dimension (UED) model \cite{Appelquist:2000nn} with a compactified
electroweak-scale 
extra space dimension, each SM particle is paired with a tower of Kaluza-Klein (KK)
states with identical spin. Thus, model-independent spin measurements are
crucial in discriminating among many extensions of the SM. \\

There have been several proposals for measuring spin at both the LHC and the prospective
$e^+e^-$ International Linear Collider (ILC) \cite{ILC}.
Threshold scans in $e^+e^-$ collisions
can be used to distinguish scalars from spinors at the ILC, as the scalar production
cross section increases slowly $\sim \beta^3$ while the spinor cross section
increases steeply $\sim \beta$ \cite{Battaglia:2005zf}.
[Such a method cannot be used at the LHC as the center of mass (c.m.) energy
at the parton level is not fixed.] 
The production angle can give insight on spin as well. The polar-angle
distribution of a $s$-channel pair produced scalars is proportional to
$\sin^2 \Theta$, while for spinors it approaches $1+\cos^2\Theta$ asymptotically
at high energies. The presence of $t/u$-channel exchanges may render the production-angle
measurement of spin more demanding \cite{Battaglia:2005zf}, although it is feasible
in some cases \cite{Tsukamoto:1993gt, Choi:2006mr}. The polar-angle dependence in
decays at the ILC \cite{Choi:2006mr} and the
invariant-mass distributions in sufficiently long decay chains at the
LHC \cite{Wang:2006hk}
can also be used for spin measurements. However, these techniques
rely strongly on the final state spins and the chiral structure of
couplings.\\

In this report we study the {\it fully-correlated} azimuthal-angle distributions in
the production of a new particle-antiparticle pair in $e^+e^-$ collisions and
both of their sequential decays \cite{MurayamaTalk}. [This work is a natural
extension to the previous works \cite{Buckley:2007th,Buckley:2008pp} where the azimuthal-angle
correlation of the production and only one of the decays has been investigated.]
These distributions develop through quantum
interference between the different helicity states in a coherent sum. By extracting
this angular dependence, we can determine which helicity states contribute to the
sum, and thus the spin of the decaying particle in a model-independent way.
To be specific, we restrict ourselves to production of an electrically-charged
particle-antiparticle ($F^+F^-$) pair in $e^+e^-$ collisions and
decay of each produced particle $F^\pm$ to a charged particle $f^\pm$ and an
invisible particle $\chi$,
\begin{eqnarray}
e^+e^-\to F^+ F^- \to (f^+ \chi ) (f^- \chi) \to f^+f^- \not\!\!{E}.
\label{eq:characteristic_process}
\end{eqnarray}
As suggested by the WIMP solution to the cold dark matter puzzle, large missing
energy signatures are considered likely at the TeV scale and are generic in
many extensions to the SM.\footnote{The SM processes
$e^+e^- \to \tau^+\tau^- \to (\pi^+\bar{\nu}_\tau) (\pi^-\nu_\tau)$
and $e^+e^-\to W^+W^-\to (\ell^+\nu_\ell) (\ell^-\bar{\nu}_\ell)$
with $\ell = e$ and $\mu$ also have a large missing energy signature
carried away by invisible neutrinos.}\\

For example, the lightest supersymmetric particle (LSP), typically the lightest
neutralino $\tilde{\chi}^0_1$, in supersymmetric (SUSY) models with $R$-parity;
and the lightest KK odd particle (LKP), typically the lightest KK gauge
boson $\gamma_1$, in UED models with KK parity are stable and escape detection.
Prototype processes
in the SUSY and UED models with the same event topologies as the
process~(\ref{eq:characteristic_process}) are
\begin{eqnarray}
&& e^+e^-\to \tilde{\mu}^+_R \, \,\, \tilde{\mu}^-_R\, \to
          (\mu^+ \tilde{\chi}^0_1) (\mu^-\tilde{\chi}^0_1)
          \to  \mu^+ \mu^- \not\!\!{E},
          \label{eq:typical_SUSY_process}\\[2mm]
&& e^+e^-\to \mu^+_{R1} \mu^-_{R1} \to
          (\mu^+ \gamma_1) (\mu^-\gamma_1)
          \to \mu^+ \mu^- \not\!\!{E}.
          \label{eq:typical_UED_process}
\end{eqnarray}
Both generate the same experimental signatures $\mu^+\mu^-\not\!\!{E}$ with
large missing energy.\\
\begin{figure}[thb]
\centering
\includegraphics[width=14cm,clip]{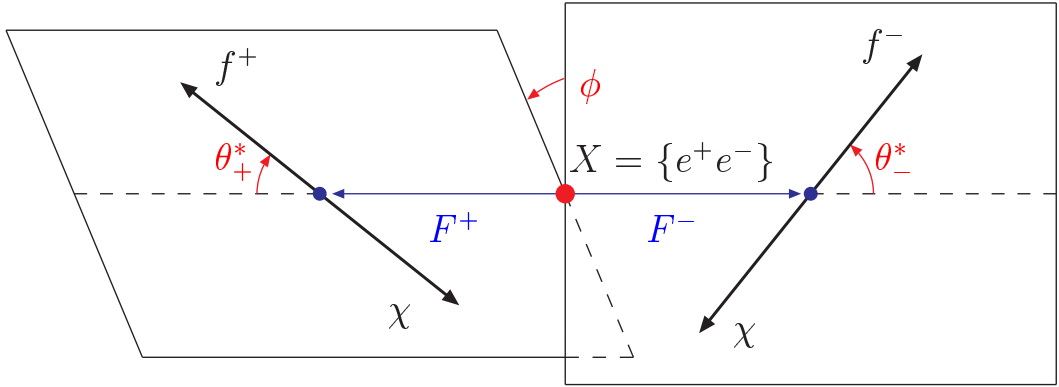} 
\caption{\it The definition of the polar angles $\theta^*_\pm$ of the visible
         particle $f^\pm$ momentum in the rest frame of the decaying particle
         $F^\pm$ and of the correlated azimuthal angle $\phi$
         between two decay planes formed in the correlated production-decay process
         $X \rightarrow F^- F^+ \rightarrow (f^-\chi) (f^+\chi)$ in the
         rest frame of the $X=\{e^+e^-\}$ system, corresponding to the $e^+e^-$
         c.m. frame for the processes considered in this report. Here,
         $X=\{e^+e^-\}$ denotes any single- or multiple-particle intermediate
         state formed in $e^+e^-$ annihilation. Note that $\phi$ is
         invariant under the Lorentz boost along the $F^\pm$ flight direction.}
\label{fig:config}
\end{figure}

The characteristic observables for measuring spin of the particles $F^\pm$
through the process (\ref{eq:characteristic_process}) are the angular distributions
of the final-state particles $f^\pm$ in the $F^\pm$ decays, encoding the helicities
of the $F^\pm$ states. We denote the polar angles of the particles $f^\pm$ in the rest
frames of the $F^\pm$ particles by $\theta^*_\pm$, and the azimuthal angles by
$\phi_\pm^*$ with respect to the production plane defined by the $e^-$ and $F^-$
momentum directions, respectively. Then the angle $\phi$ with its range $[0, 2\pi]$
between the two decay planes (see Fig.$\,$\ref{fig:config}) is the azimuthal angle
defined by the angle difference $\phi\equiv \phi^*_+ - \phi^*_-\, \, (\mbox{mod}\,\,
2 \pi$) invariant under any Lorentz boost along the $F^\pm$ flight direction.\\

If we label the $F^\pm$ helicities by $\lambda_\pm$ and $\lambda'_\pm$, the
joint production-decay distribution reads:
\begin{eqnarray}
W(E_{\rm cm} ; \Theta ; \theta^*_\pm,\phi^*_\pm)
= \sum_{\lambda_\pm, \lambda'_\pm=-j}^j
  {\cal P}^{\lambda_-\lambda_+}_{\lambda'_-\lambda'_+} (E_{\rm cm}, \Theta)\,
  {\cal D}^-_{\lambda_-\lambda'_-} (\theta^*_-,\phi^*_-)\,
  {\cal D}^+_{\lambda_+\lambda'_+} (\theta^*_+,\phi^*_+),
\label{eq:joint_distribution}
\end{eqnarray}
where $E_{\rm cm}$ is the $e^+e^-$ c.m. energy and $\Theta$ is the production
angle of $F^-$ with respect to the $e^-$ direction,
and
$j$ is the spin of the particle $F^\pm$. The production density matrix ${\cal P}$
is defined in terms of the helicity amplitudes ${\cal T}$
of the process $e^+e^-\to F^+F^-$ for unpolarized beams by
\begin{eqnarray}
{\cal P}^{\lambda_-\lambda_+}_{\lambda'_-\lambda'_+}
 =
 \sum_{\sigma_\pm=\pm 1/2}
 {\cal T}_{\sigma_-\sigma_+;\lambda_-\lambda_+}
 {\cal T}^*_{\sigma_-\sigma_+;\lambda'_-\lambda'_+},
\label{eq:production_density_matrix}
\end{eqnarray}
where $\sigma_\pm$ is the $e^\pm$ helicity, and each $F^\pm$ decay density matrix
${\cal D}^\pm$ has a simple azimuthal-angle dependence of a pure kinematical
origin as
\begin{eqnarray}
{\cal D}^\pm_{\lambda_\pm \lambda'_\pm} (\theta^*_\pm, \phi^*_\pm)
 =
 D^\pm_{\lambda_\pm \lambda'_\pm}(\theta^*_\pm)\,\,
 {\rm e}^{\mp i(\lambda_\pm-\lambda'_\pm)\phi^*_\pm},
\label{eq:decay_density_matrix}
\end{eqnarray}
reflecting an overall rigid rotation of the decay plane around the parent
particle momentum.\\

Integrating the joint production-decay distribution $W$ in
Eq.$\,$(\ref{eq:joint_distribution}) over the production angle $\Theta$, the
decay angles $\theta^*_\pm$ and $\phi^*_+$ with the azimuthal angle
$\phi$ fixed, we can derive the correlated azimuthal-angle distribution between
the two decay planes as
\begin{eqnarray}
\frac{d{\cal C}}{d\phi}
  = \int W(E_{\rm cm} ; \Theta ; \theta^*_\pm, \phi^*_\pm)\,\, d\cos\Theta\,
                      d\cos\theta^*_- d\cos\theta^*_+\, d\phi^*_+.
\end{eqnarray}
We note from Eqs.$\,$(\ref{eq:joint_distribution}) and (\ref{eq:decay_density_matrix})
that the dependence on $\phi^*_+$ is of the form
${\rm exp}[-i\phi^*_+ (\lambda_--\lambda'_- -\lambda_+ + \lambda'_+)]$ so that the
integral over $\phi^*_+$ leaves only those terms in
Eq.$\,$(\ref{eq:joint_distribution}) satisfying the relation $\lambda_+ - \lambda'_+
=\lambda_--\lambda'_-\equiv\Lambda $ in  the range $[-2j, 2j]$.
If the distribution is further integrated over the angle $\phi$, only the
incoherent terms with $\lambda_\pm=\lambda'_\pm$ survive. Thus,
any non-trivial azimuthal-angle distribution indicates the presence of
quantum interference between the different helicity amplitudes. \\

In weakly-interacting and CP-invariant theories with negligible particle-width and
loop effects, the general form of the normalized azimuthal-angle correlation for the
production and decay of a spin-$j$ particle pair\footnote{Even in the
CP-noninvariant case, all the sine terms are washed out by taking the average over
two possible azimuthal angles, which is unavoidable
due to a two-fold ambiguity in reconstructing the $F^\pm$ momentum
as described in the following.} is
\begin{eqnarray}
\frac{1}{\cal C}\frac{d{\cal C}}{d\phi}
  = \frac{1}{2\pi}
    \left[1 + A_1 \cos(\phi) + \cdots +  A_{2j} \cos (2j \phi)\right].
\label{eq:cos_phi_distribution}
\end{eqnarray}
Each coefficient can be worked out from the standard rules of constructing matrix
elements. However, it is guaranteed on a general footing  \cite{Choi:2006mr} that
the highest non-vanishing coefficient is always $A_{2j}$. This is because the
production
of a charged pair $F^\pm$ in $e^+e^-$ collisions gets a non-zero spin-1
photon-exchange contribution to the production amplitudes, ${\cal T}_{\sigma_-
\sigma_+;\pm j \pm j}$. However this term tends to be suppressed by
$\sim m^2_{F^\pm}/E^2_{\rm cm}$
at high energies because of a final-state helicity flip.
Thus, the spin $j$ can
be determined by identifying the highest $\cos(2j\phi)$ mode at a c.m. energy
not far away from the production threshold, if the production amplitudes contributing
to the coefficient $A_{2j}$ are not so suppressed and the decays $F^\pm \to
f^\pm \chi$ do not have too small polarization analyzing powers.\\

The correlated azimuthal angle distribution for the smuon-pair
process (\ref{eq:typical_SUSY_process}) is {\it flat, as it must be,} and
the distribution for the KK muon-pair process (\ref{eq:typical_UED_process})
is given by
\begin{eqnarray}
\frac{1}{\cal C} \frac{d{\cal C}}{d\phi} \left[\mu^+_{R1}\mu^-_{R1}\right]
 =
 \frac{1}{2\pi} \left[\, 1 - \frac{\pi^2\, m^2_{\mu^\pm_{R1}}}{8(E_{\rm cm}^2
                       + 2 m^2_{\mu^\pm_{R1}})}
                   \left(\frac{1-2 m^2_{\gamma_1}/m^2_{\mu^\pm_{R1}}}{
                               1+ 2 m^2_{\gamma_1}/m^2_{\mu^\pm_{R1}}}\right)^2
                   \cos\phi\, \right].
\label{eq:kk_muon_correlation}
\end{eqnarray}
It is apparent from the expression (\ref{eq:kk_muon_correlation}) that (a) the
coefficient of the highest $\cos\phi$ mode is maximal in magnitude at the production
threshold and it decreases rapidly with increasing energy in conformity to the
general rule as outlined above,
and (b) it is very sensitive to the values of the $\mu_{R1}$ and
$\gamma_1$ masses, leading to the restriction that the magnitude of the coefficient
$A_1$ of the highest $\cos\phi$ mode cannot be larger than $\pi^2/48\simeq 0.206$. \\

\begin{figure}[thb]
\centering
\includegraphics[width=9cm]{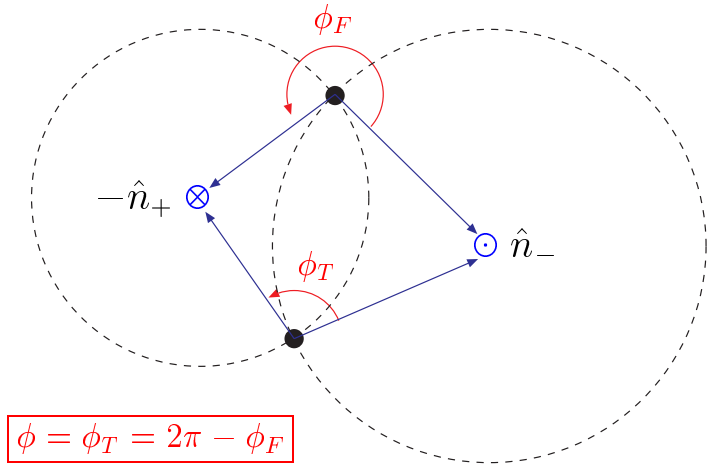}
\caption{\it An illustration for the relation $\phi_F=2\pi-\phi_T$ between the
         true and false azimuthal angle differences, $\phi_T=\phi$ and $\phi_F$,
         leading to the equality: $\cos\phi_F=\cos\phi_T = \cos\phi$.
         The dashed-line circles are the circles projected on a sphere of a unit
         radius (centered on the $e^+e^-$ interaction) of two cones satisfying the
         relation (\ref{eq:cone_relation}).
         The solid black dots indicate
         two solutions for the $F^\pm$ direction and the unit vector $\hat{n}_\pm$
         stands for the $f^\pm$ direction. }
\label{fig:true_false}
\end{figure}

As well known \cite{Tsukamoto:1993gt,Choi:2006mr}, there exists a two-fold discrete
ambiguity in completely reconstructing the $F^\pm$ four-momenta and thus the azimuthal
angles $\phi_\pm$ in the process (\ref{eq:characteristic_process}) in the
laboratory frame, even if all particle masses are known. In \cite{Buckley:2007th} it was shown that this ambiguity could obscure the helicity information in $\phi^\pm$, curtailing their use as measurements of spin. 
{\it Nevertheless, the cosine
of the azimuthal angle $\phi$ is unambiguously determined by measuring the $f^\pm$
four-momenta event by event.} To prove this important point analytically, let
the pair produced  particles $F^\pm$ and the invisible particle $\chi$ have mass
$m_\pm$ and $m_0$ and denote the $f^\pm$ flight direction in the laboratory
frame by a unit vector $\hat{n}_\pm$, respectively. Then, the opening angles
$\theta_\pm$ between the visible $f^\pm$ tracks and the parent $F^\pm$
particles in the laboratory frame can be determined from the relation
\begin{eqnarray}
  m^2_\pm - m^2_0
= E_{\rm cm}\, E_{f^\pm}
  \left(1-\sqrt{1-4 m^2_\pm/E^2_{\rm cm}} \cos\theta_\pm\right),
\label{eq:cone_relation}
\end{eqnarray}
defining two cones about the $f^+$ and $f^-$ axes which
intersect in two lines - the true $F^\pm$ flight direction and a false direction.
True and false solutions are mirrored on the plane spanned by the $f^+$
and $f^-$ flight directions, leading to the relation
$\phi_T = 2\pi -\phi_F =\phi$
between the true and false values, $\phi_T$ and $\phi_F$, of the
azimuthal angle $\phi$ (see Fig.$\,$\ref{fig:true_false}).
Therefore, the cosine of the
azimuthal angle is uniquely determined and its expression is given by
the simple expression:
\begin{eqnarray}
\cos\phi = (\hat{n}_+\cdot\hat{n}_- + \cos\theta_+ \cos\theta_-)/
            \sin\theta_+\sin\theta_-,
\end{eqnarray}
expressed in terms of the unit vectors, $\hat{n}_\pm$, and the opening
angles, $\theta_\pm$.\\

Two cosines, $\cos (n_a\phi)$ and $\cos (n_b\phi)$, for any integers
$n_a$ and $n_b$, are functions of $\cos\phi$ and orthogonal to each other
for $n_a\neq n_b$. Therefore, we can project out all the coefficients $A_k$
($k=1,\ldots, 2j$) by fitting the expression of $(1/{\cal C})\, d{\cal C}/d\phi$
in Eq.$\,$(\ref{eq:cos_phi_distribution}) to the distribution measured
experimentally.\\

For a numerical demonstration of this spin-determination method, we compare the
correlated azimuthal-angle $\phi$ distribution of the SUSY process
(\ref{eq:typical_SUSY_process}) with that of the UED process
(\ref{eq:typical_UED_process}) in a specific scenario, which we will simply
call ``BCMM" for convenience in the following, with the particle mass spectrum,
\begin{eqnarray}
\mbox{\underline{BCMM}} :\,
m_\pm = m_{\tilde{\mu}^\pm_R} =m_{\mu^\pm_{R1}} = 200\, \mbox{GeV}\ \
\mbox{and}\ \
m_0 = m_{\tilde{\chi}^0_1} = m_{\gamma_1} = 50\, \mbox{GeV}.
\end{eqnarray}
We stress that the mass spectrum is chosen only as a simple illustrative
example for SUSY and UED models with different spins but similar final
states and so the spin-determination method demonstrated here can, in
principle, be exploited equally for any other scenarios beyond as well as within
the SM.\\

\begin{figure}[htb]
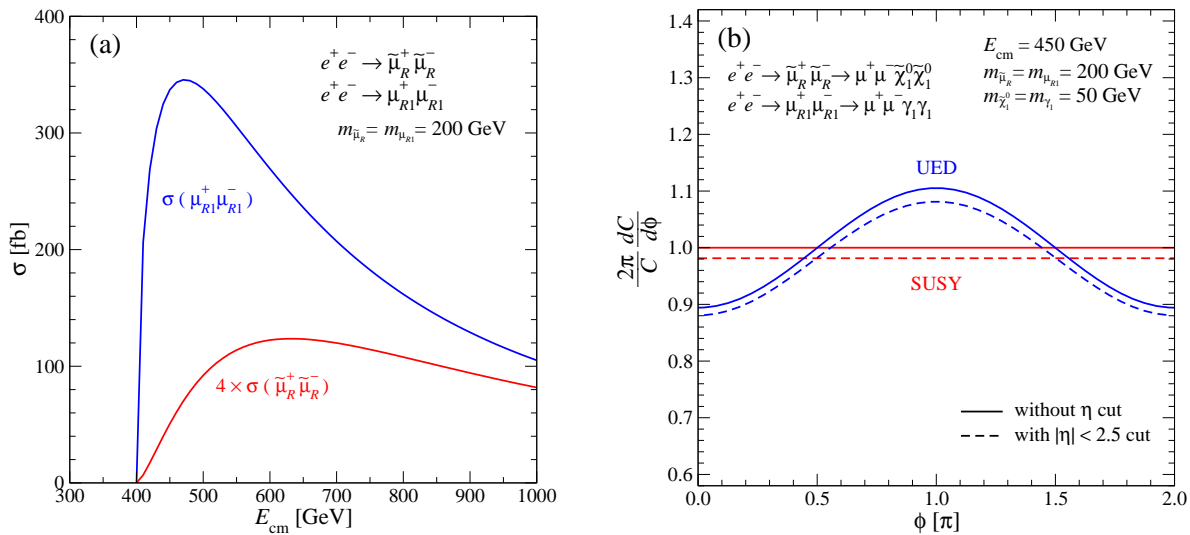

\centering
\includegraphics[height=7cm,clip]{xsec_smuon_kk_muon.eps}\qquad
\includegraphics[height=7cm,clip]{phi_smuon_kk_muon.eps}
\caption{\it (a) The c.m. energy dependence of the production cross sections for
             smuons and KK muons; and (b) the normalized azimuthal-angle
             distribution $(2\pi/{\cal C}) d{\cal C}/d\phi$ in the BCMM scenario
             with $E_{\rm cm}=450$ GeV. The solid (dashed) line is the distribution
             without (with) the rapidity cut on the $\mu^\pm$ directions
             and the total missing momentum, $|\eta|< 2.5$.
\label{fig:smuon_kk_muon_dist}}
\end{figure}

The total c.m. energy at the ILC is expected to reach up
to $1~\mbox{TeV}$, and an integrated luminosity of $500~\mbox{fb}^{-1}$
is not unrealistic. For the mass spectra chosen, we expect several
thousand to several hundreds of thousands of events available as shown
in Fig.$\,$\ref{fig:smuon_kk_muon_dist}(a). [At high energies we note that the $\mu^\pm_{R1}$ scalar
cross section scales in the same way as the $\tilde{\mu}^\pm_R$ spinor cross
section, but with a coefficient 4 times as
large due to differences in the number of spin degrees of freedom, as
familiar from QED processes.] \\

To simulate the effects of experimental cuts which are unavoidable due to the geometry
of the detector, we place cuts on the pseudo-rapidity of the $\mu^\pm$ directions and
the total missing momentum: $|\eta| <2.5$, as otherwise the leptons
would vanish unseen down the beam and the missing momentum is not guaranteed to
be carried away by the invisible $\chi$ particles.
Two representative distributions for scalar and spinors (both with and without rapidity cuts)
are shown in Fig.~\ref{fig:smuon_kk_muon_dist}(b).  The SUSY distribution is flat
and the UED distribution has a clear $\cos\phi$ dependence above a flat
distribution, as expected. Furthermore, the rapidity cut reduces  the total
number of events by about 2\% but hardly modifies the correlated azimuthal-angle
distribution. It is therefore apparent even at this level of analysis that the
non-trivial correlated azimuthal-angle distribution contains the spin-1/2
information of the KK muon, $\mu^\pm_{R1}$.\\ 

\begin{figure}[htb]
\centering
\includegraphics[height=8cm,clip]{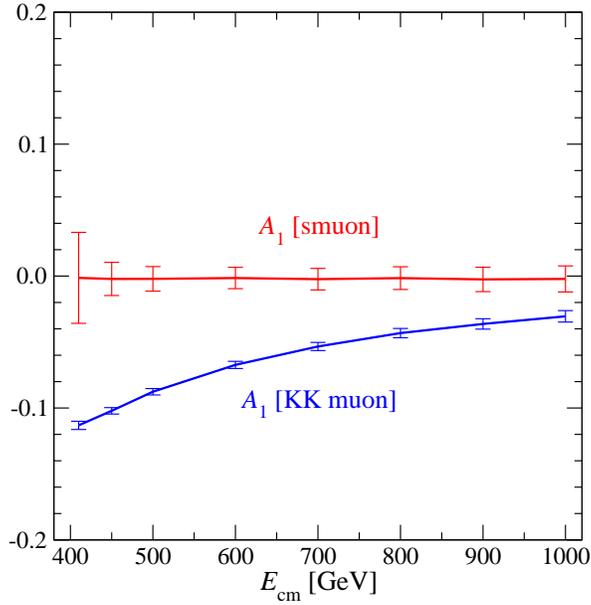}
\caption{\it Coefficient $A_1$ for the BCMM point as a function of the $e^+e^-$
             c.m. energy for both the SUSY $\tilde{\mu}^\pm_R$ and
             UED $\mu^\pm_{R1}$ pair production with an integrated luminosity of
             $500~\mbox{fb}^{-1}$. Error bars obtained with $Br(\tilde{\mu}^\pm_R
             \to\mu^\pm\tilde{\chi}^0_1) = Br(\mu^\pm_{R1}\to\mu^\pm \gamma_1)=1$
             correspond to the 1-$\sigma$ uncertainty range.
\label{fig:a1a0}}
\end{figure}

Using the least-square method we fit the generated distributions to
$(1+A_1\cos\phi+A_2\cos 2\phi)/2\pi$ after placing the cut on the pseudo-rapidities of
the $\ell^\pm$ and the total missing momentum. 
Only the coefficient
$A_1$ for the scalar $\tilde{\mu}_R$ and spinor $\mu_{R1}$ are shown in
Fig.~\ref{fig:a1a0}, because the coefficient $A_2$ is found to be extremely
small ($< 0.1\%$). This is consistent with the statistical errors introduced
from the event generation and fitting procedure and confirms our general
argument that $A_i = 0$ for $i > 1$.
As can be seen, in the BCMM spectrum
the values of $A_1$ for the scalar muons, $\tilde{\mu}_R$, are consistent with zero
for all energies, as expected. For the KK muons, $\mu^\pm_{R1}$, the coefficient
$A_1$ is manifestly non-zero, allowing us to clearly
distinguish the spinor field from spin-0 scalar states.\\

Due to the large $A_1$ signal for the $\mu^\pm_{R1}$ process on the order of $10\%$
and the negligible rapidity-cut effect on the normalized azimuthal-angle distribution,
the ILC with a large integrated luminosity can discern that the particle is not
a spin-0 but a spin-1/2 particle. However it is not always guaranteed that, in the
case of
higher spins, the highest mode is sufficiently large to be discerned. As a result,
the question still remains whether this spin-determination method may be practically
applied to discriminate spinors from vectors and vice versa in general cases.
It is worthwhile to study the case of pair production of massive spin-1/2
fermions in SUSY contrasted with vector production in UED.\\

In particular, we consider the processes - the production and two-body decays of
a lighter chargino pair and those of a first KK $W$-boson pair:
\begin{eqnarray}
&& e^+e^- \to\, \tilde{\chi}_1^+\,\, \tilde{\chi}_1^-\,\, \to
   (\ell^+\, \tilde{\nu}_{\ell})\,  (\ell^- \, \tilde{\nu}^*_{\ell}), \\
&& e^+e^- \to W_1^+W_1^- \to
   (\ell^+\nu_{\ell 1})(\ell^- \bar{\nu}_{\ell 1}),
\end{eqnarray}
where the charged leptons $\ell^\pm$ can be of either a muon or an electron type.
An explicit calculation shows that, when CP is preserved and all the absorptive
parts are negligible, the correlated azimuthal-angle
distributions $(2\pi/{\cal C})\, d{\cal C}/d\phi$ are given in terms of the
production matrices by
\begin{eqnarray}
 &&  \mbox{\fbox{$\,\tilde{\chi}^+_1\,\,\, \tilde{\chi}^-_1$}}
      \, :\,
      1 + \frac{\pi^2}{8}\, \Re {\rm e}
      \, ( \rho^{++}_{--})\, \cos(\phi), \\[3mm]
 &&  \mbox{\fbox{\small $W^+_1W^-_1$}} \, :
      \,  1-\frac{9\pi^2}{32}
      \left( \frac{2 m^2_{W^\pm_1}}{2m^2_{W^\pm_1} + m^2_{\nu_1}}\right)^2
      \Re {\rm e}\, ( \rho^{++}_{\, 0\, 0}
             + \rho^{\,0\, 0}_{--}
             + \rho^{+ 0}_{\, 0-}
             + \rho^{\,0 +}_{- 0})\,
       \cos(\phi)  \nonumber \\[2mm]
 &&  \mbox{ }\hskip 3.1cm
      +\left(\frac{m^2_{W^\pm_1}-m^2_{\nu_1}}{2m^2_{W^\pm_1}+m^2_{\nu_1}}\right)^2
       \Re {\rm e}\, (\rho^{++}_{--} +\rho^{--}_{++} )\, \cos (2\phi),
\label{eq:azimuthal_distribution_chi1_w1}
\end{eqnarray}
where the super/sub-scripts $\pm$ stand for the helicities $\pm 1/2$ and $\pm 1$ of
$\tilde{\chi}^\pm_1$ and $W^\pm_1$, respectively.  The integrated production density
matrix $\rho$ is defined as:\footnote{A detailed derivation
of the explicit forms of the production density matrices will be presented in a
separate publication.}
\begin{eqnarray}
\rho^{\lambda_-\lambda_+}_{\lambda'_-\lambda'_+}
  =
\int {\cal P}^{\lambda_-\lambda_+}_{\lambda'_-\lambda'_+}(E_{\rm cm},\Theta)
     \, d \cos\Theta \mbox{\Large $/$}
     \sum_{\lambda_\pm}
\int {\cal P}^{\lambda_-\lambda_+}_{\lambda_-\lambda_+}(E_{\rm cm}, \Theta)\,
    d \cos\Theta.
\end{eqnarray}
We note that the $\tilde{\chi}^\pm_1$ and $W^\pm_1$ two-body decays do not suppress
the $\cos\phi$ terms, while the highest $\cos(2\phi)$ mode in the $W^\pm_1$ case may
be strongly suppressed due to a small polarization analyzing power of the $W^\pm_1$
decay.
In particular, the coefficient becomes vanishing if two states, $W^\pm_1$ and
$\nu_1$, are degenerate. \\

For our numerical demonstration in the $\tilde{\chi}^\pm_1$ and $W^\pm_1$ cases, we
take the masses of $\tilde{\chi}^\pm_1/W^\pm_1$ and $\tilde{\nu}_\ell/\nu_{\ell 1}$
in our BCMM scenario to be
\begin{eqnarray}
\mbox{\underline{BCMM}}: \,
m_\pm = m_{\tilde{\chi}_1^\pm}=m_{W_1^\pm} = 300~\mbox{GeV}
\ \
{\rm and}
\ \
m_0 = m_{\tilde{\nu}_\ell}=m_{\nu_{\ell 1}}= 200~\mbox{GeV}.
\end{eqnarray}
Even though the $\nu_{\ell 1}/\tilde{\nu_\ell}$ are not the LSP/LKP, they decay to
neutrinos and the LSP/LKP, neither of which is visible in the detector.\\

\begin{figure}[htb]
\centering
\includegraphics[height=8cm,clip]{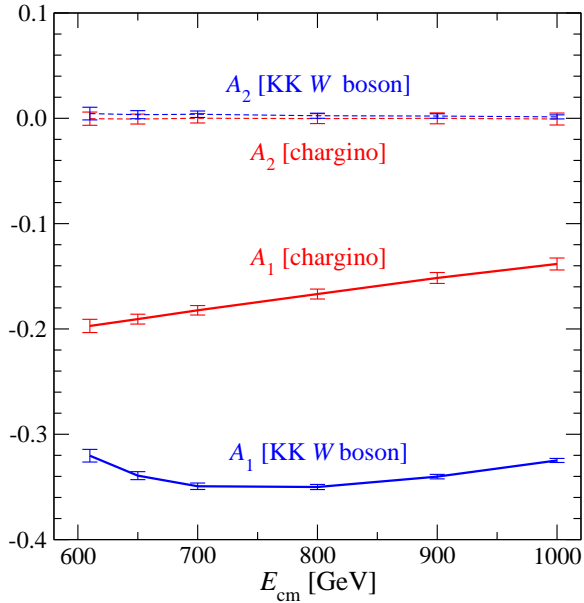}
\caption{\it Coefficients $A_1$ and $A_2$ as a function of $E_{\rm cm}$
             for the SUSY  $\tilde{\chi}^\pm_1$ and the UED $W^\pm_1$
             production in the BCMM spectrum with an integrated luminosity of
             $500~\mbox{fb}^{-1}$. Error bars obtained with
             $Br(\tilde{\chi}^\pm_1\to \ell^\pm\tilde{\nu}_{\ell})
              =Br(W^\pm_1\to \ell^\pm\nu_{\ell 1})=0.4$
             correspond to the 1-$\sigma$ uncertainty range.
\label{fig:WSPS3}}
\end{figure}

We perform fits of the azimuthal angle distributions obtained with an integrated
luminosity of $500~\mbox{fb}^{-1}$ to $(1+A_1\cos\phi+A_2\cos2\phi)/2\pi$. The
results for the BCMM mass spectrum are displayed in Fig.~\ref{fig:WSPS3}.
Error bars corresponding to the 1-$\sigma$ uncertainty range are
obtained with $\mbox{Br}(\tilde{\chi}^\pm_1\to\ell^\pm\tilde{\nu}_{\ell})
=\mbox{Br}(W^\pm_1\to \ell^\pm\nu_{\ell 1})=0.4$.
First, we note that the coefficient $A_2$ is too small (less than 0.5\%)
in magnitude to distinguish the spin-1 $W^\pm_1$ state from the spin-1/2
$\tilde{\chi}^\pm_1$ state. This strong suppression in the $W^\pm_1$ case is
not only due to the small analyzing power
 of the $W^\pm_1$ two-body decays
in the BCMM scenario but also due to the cancellation of the corresponding
production helicity amplitudes, $\sim m^2_{W^\pm_1}/E^2_{\rm cm}$, that is
forced by electroweak gauge invariance to save the unitarity \cite{Duncan:1985ij}.
On the contrary, the coefficient $A_1$ in both the $\tilde{\chi}^\pm_1$ and the
$W^\pm_1$ case is sufficiently large so that they can clearly be distinguished
from the spin-0 case.\\

{\it To conclude}. The fully-correlated azimuthal-angle correlations encoding quantum
interference between different helicity states can provide a powerful method of spin
measurements at the ILC. We have found that, if all the particle masses
are known, all the cosines of the azimuthal angle between two decay planes in the
process (\ref{eq:characteristic_process}) are unambiguously determined despite
the inherent 2-fold discrete ambiguity in determining the four-momenta of the
decaying particles. \\

Quantitatively, we have shown in a specific scenario that the spin-0 smuons
can be distinguished from spin-1/2 KK muons or higher-spin states. However, it
turned out to be difficult to establish the spin-1 nature of the KK $W$-boson
due to the strong suppression of the highest $\cos 2\phi$ mode, requiring
other methods such as the decay polar-angle distributions \cite{Choi:2006mr} and the
singly-correlated azimuthal-angle distributions \cite{Buckley:2007th}.
However, the latter suffers from the two-fold ambiguities in the reconstruction.\\

Before closing, we emphasize once more that, although applied only to the SUSY and
UED processes explicitly in this report, the proposed spin-determination method
through quantum azimuthal-angle correlations can be used for any process with a
generic event topology similar to that in the process (\ref{eq:characteristic_process})
in the SM and beyond.\\

\vskip 0.8cm

\subsection*{Acknowledgments}

KM acknowledges the hospitality of the Chonbuk National University where
part of this work was carried out.
The work of SYC was supported in part by the Korea Research Foundation Grant
funded by the Korean Government (MOERHRD, Basic Research Promotion Fund)
(KRF-2007-521-C00065) and in part by KOSEF through CHEP at Kyungpook
National University. The work of MRB and HM was supported in part by World
Premier International Research Center Initiative (WPI Initiative),
MEXT, Japan, in part by the U.S. DOE under Contract DE-AC03-76SF00098,
and in part by the NSF under grant PHY-04-57315.

\end{document}